\documentclass[onecolumn, draftclsnofoot]{IEEEtran}

\usepackage{amsmath, amssymb}

\newtheorem{theorem}{Theorem}

\newtheorem{lemma}{Lemma}

\begin{document}

\title{Single machine slack due-window assignment and scheduling of linear time-dependent deteriorating jobs and a deteriorating maintenance activity}

\author{\authorblockN{Bo Cheng\authorrefmark{1} and Ling Cheng\authorrefmark{2}\\ 
}
\authorblockA{\authorrefmark{1}School of Informatics\\Guangdong University of Foreign Studies\\Guangzhou 510420, China\\
Email: bocheng@gdufs.edu.cn\\}
\authorblockA{\authorrefmark{2}School of Electrical and Information Engineering\\University of the Witwatersrand\\Private Bag 3, Wits. 2050, Johannesburg, South Africa\\
Email: ling.cheng@wits.ac.za}}

\maketitle

\begin{abstract}
In this paper, we consider the slack due-window assignment model and study a single machine scheduling problem of linear time-dependent deteriorating jobs and a deteriorating maintenance activity. The cost for each job consists of four components: earliness, tardiness, window location and window size. The objective is to schedule the jobs and to assign the maintenance activity and due-windows such that the total cost among all the jobs is minimized. A polynomial-time algorithm with the running time not exceeding $O(n^2logn)$ to give a solution to this problem is introduced, where $n$ is the number of jobs. 

\end{abstract}



\section{Introduction}
\label{sec1}

Competition in market place prompts the studies on operations management to improve customer service. One important objective of operations management in practice is to finish jobs as close as possible to their due-dates. Usually a time interval is assigned in the supply contract so that a job completed within the time interval will be considered on time and not be penalized. The time interval is called the due-window of a job. The due-window assignment methods include common due-window, slack due-window (also called common flow allowance) and others. Some relevant references are \cite{cheng1988optimal, liman1998common, yeung2001minimizing, yeung2001single,  yeung2004two, yeung2009two, yeung2010optimal, gerstl2013due, gerstl2013improved, yang2014scheduling, yin2013single}.  \cite{mosheiov2008multi} presented a polynomial-time solution to find the optimal schedule and the optimal due-window size such that the total cost is minimized. \cite{mosheiov2010job} considered the slack due-window problem and gave an $O(nlogn)$ time complexity solution.

In classical scheduling theory, job processing times are considered as constants. However, a steadily growing interest on solving scheduling problems with changeable job processing times has been witnessed in the last decade. Recently, \cite{wang2011single} studied the single machine common due-window assignment problem with learning effect and deteriorating jobs. They gave polynomial-time algorithms to minimize costs for earliness, tardiness, window location and window size. \cite{wang2013single} considered the parallel problem for the slack due-window model. 

\cite{lee2001machine} initiated research on machine scheduling with a rate-modifying activity. Since then, researchers have applied the concept of rate-modifying activity to other scheduling settings involving various performance measures (\cite{mosheiov2006due,  mosheiov2009schedulingB, mosheiov2010scheduling,  ji2010scheduling}). In this paper, the maintenance activity considered is different from the rate-modifying activity. It can be described as follows. Assume that there is at most one maintenance activity throughout the schedule. Maintenance activity can be performed immediately after the completion of any job. However, the position and starting time of the maintenance activity are decided by the scheduler. The machine reverts to its initial conditions after the maintenance activity including machine deterioration. See \cite{yang2010single, zhao2010single} and \cite{cheng2012common}.

The combinations of the above-mentioned settings have been considered in the following recent literatures. Common due-window assignment and scheduling with simultaneous considerations of time-dependent deteriorating jobs and a rate-modifying activity was studied in \cite{zhao2012note}. Slack due-window assignment and scheduling with considerations of variable processing-time jobs and a rate-modifying activity was considered in \cite{mor2012scheduling}. Common due-window assignment and scheduling with simultaneous considerations of time-dependent deteriorating jobs and a maintenance activity was investigated in \cite{cheng2012common}. In this paper, the problem of slack due-window assignment and single-machine scheduling considering time-dependent deteriorating jobs and a maintenance activity is presented. To our best knowledge, this problem has not been studied in literatures.

The rest of this paper is organized as follows. In Section~\ref{sec2}, a description of the problem is given. In Section~\ref{sec3} some important lemmas and properties are presented. In Section~\ref{sec4}, a polynomial-time solution for the problem is given. A numerical example is presented to demonstrate the polynomial-time solution in Section~\ref{sec5}. The research is concluded and future study is foreseen in the last section. 

\section{Model formulation} 
\label{sec2}

There are $n$ jobs $J_1, J_2, \dots, J_n$ to be processed on a single machine. All the jobs are available for processing at time zero and  no preemption is allowed. The job processing times are assumed to follow a linear time-dependent deteriorating model. Then, the actual processing time of job $J_j$ is determined by
\begin{equation}
p_j=a_j+bt, j=1,2,\ldots, n,
\end{equation}
where $a_j$ is the normal processing time of job $J_j$, $b$ is a common deteriorating factor for all the jobs, and $t\geq 0$ is the starting time of job $J_j$.

The due window of job $J_j$ is specified by a pair of non-negative real numbers $[d_j^{(1)}, d_j^{(2)}]$ such that $d_j^{(1)} \leq d_j^{(2)}$. For a given schedule $\pi$, $C_j=C_j(\pi)$ denotes the completion time of job $J_j$, $E_j ={max} \{0, d_j^{(1)}-C_j\}$ is the earliness value of job $J_j$, $T_j = {max} \{0, C_j-d_j^{(2)}\}$ is the tardiness value of job $J_j$, and $D_j=d_j^{(2)}-d_j^{(1)}$ is the due-window size of job $J_j$. For the slack due-window method, the window starting time for job $J_j$ is defined as the sum of its processing time $p_j$ and a job-independent constant $q^{(1)}$:
\begin{equation}
d_j^{(1)}= p_j+q^{(1)},
\end{equation}
and the due window completion time for $J_j$ is defined as the sum of its processing time and a job-independent constant $q^{(2)} > q^{(1)}$:
\begin{equation}
d_j^{(2)}= p_j+q^{(2)}.
\end{equation}

Then $D_j=q^{(2)}-q^{(1)}$, for $j=1,\dots,n$, i.e., the window size is identical for all the jobs. Let $D=D_j$.

Furthermore, the following assumptions have been made for this problem: (i) the machine will revert to its initial conditions after the maintenance activity and machine deterioration will start anew, (ii) there is at most one maintenance activity throughout the schedule, and (iii) the maintenance duration is a linear function of its starting time and is given by $f(t)=\mu +\sigma t$, where $\mu >0$ is the basic maintenance time, $\sigma$ is a maintenance factor, and $t$ is the starting time of the maintenance activity. 

The objective function consists of four cost components, i.e. (i) earliness $E_j$, (ii) tardiness $T_j$, (iii) the starting time of the due-window $d_j^{(1)}$, and (iv) the due-window size $D$. 
Let $\alpha >0$, $\beta >0$, $\gamma >0$ and $\delta >0$ represent the earliness, tardiness, due-window starting time and due-window size costs per unit time respectively.  The general objective is to determine the optimal $q^{(1)}$ and $q^{(2)}$, the optimal location of the maintenance activity, and the optimal schedule to minimize the total cost function

\begin{equation}
\label{eq:z}
Z=\sum_{j=1}^n(\alpha E_j+\beta T_j+\gamma d_j^{(1)}+\delta D).
\end{equation}

Using the three-field notation of \cite{graham1979optimization}, the problem under study is denoted as $1\mid SLK, p_j=a_j+bt, ma\mid \sum_{j=1}^n(\alpha E_j+\beta T_j+\gamma d_j^{(1)}+\delta D)$, where $SLK$ and $ma$ in the second field denote the slack due-window method and maintenance activity, respectively.

For convenience, we define $\sum_{j=s}^{t} x_j=0$ if $t<s$.

\section{Properties of an optimal solution}
\label{sec3}

In this section some properties for an optimal schedule are obtained. 

\begin{lemma}
\label{le1}
If $C_j \geq d_j^{(2)}$ for a given job order $\pi =(J_1, J_2, \dots, J_n)$, then $C_{j+1} \geq d_{j+1}^{(2)}$. 
\end{lemma}

{\bf{Proof:}} We have 
\begin{eqnarray*}
	&C_{j+1}	\geq C_j+p_{j+1} \geq d_j^{(2)}+p_{j+1}        \\
 					&=q^{(2)}+p_j+p_{j+1} =d_{j+1}^{(2)}+p_j \geq d_{j+1}^{(2)}  					
\end{eqnarray*} \hfill $\Box$
 
Similar to the proof of Lemma \ref{le1}, we have

\begin{lemma}
\label{le2}
If $C_j \leq d_j^{(1)}$ for a given job order $\pi =(J_1, J_2, \dots, J_n)$, then $C_{j-1} \leq d_{j-1}^{(1)}$. 
\end{lemma}

Consider a job sequence $\pi =(J_1, J_2, \dots, J_n)$. Assume that $C_s\leq q^{(1)}\leq C_{s+1}$ and $C_t\leq q^{(2)}\leq C_{t+1}$.
Then the total cost $Z$ is a linear function of $q^{(1)}$ and $q^{(2)}$, and thus an optimum is obtained either at $q^{(1)}=C_s$ or $q^{(1)}=C_{s+1}$ and either at $q^{(2)}=C_t$ or $q^{(2)}=C_{t+1}$.

Therefore we obtain the following result, whose proof is similar to the one in \cite{mosheiov2010job}.

\begin{lemma}
\label{le3}
(i) For any given schedule, the optimal values of $q^{(1)}$ and $q^{(2)}$ are determined by the completion times of the $k$'th and $l$'th jobs ($l\geq k$).
 
(ii) An optimal schedule exists with no idle time between consecutive jobs and starts at time zero.
\end{lemma}

For a number $a$, $\left\lfloor a \right\rfloor$ denotes the largest integer not more than $a$.

\begin{lemma}
\label{le4}
$k=\left\lfloor \frac{n(\delta-\gamma)}{\alpha} \right\rfloor$ and $l=\left\lfloor \frac{n(\beta-\delta)}{\beta}\right\rfloor$
\end{lemma}

{\bf{Proof:}} Shift $q^{(1)}$ to the left by $\Delta$  time units, where $0<\Delta<p_k$. As a result, the overall cost $Z$ has been changed by $(\delta n-n\gamma -\alpha k) \Delta $. Since $q^{(1)}=p_1+p_2+\dots +p_k$ is optimal, it implies that $(\delta n-n\gamma -\alpha k) \Delta \geq 0$, and hence $k\leq \frac{n(\delta - \gamma)}{\alpha}$.  

Shift $q^{(1)}$ to the right by $\Delta$ time units, where $0<\Delta<p_{k+1}$. As a result, the overall cost $Z$ has been changed by $(\alpha (k+1)+n\gamma -\delta n ) \Delta $. We obtain $(\alpha (k+1)+n\gamma -\delta n ) \Delta \geq 0$, and hence $k\geq \frac{n(\delta - \gamma)}{\alpha}-1$. Then  $k=\left\lfloor \frac{n(\delta-\gamma)}{\alpha} \right\rfloor$. 

In the similar way, we can prove $l=\left\lfloor \frac{n(\beta-\delta)}{\beta}\right\rfloor$ by using the standard perturbation method.   \hfill $\Box$

\begin{lemma}
\label{le5}

Suppose that sequences $x_1, x_2, \ldots , x_n$ and $y_1, y_2, \ldots , y_n$ are given except in arrangement. The sum of the products of the corresponding elements  $\sum_{j=1}^n x_j y_j$ is minimized if the sequences are monotonic in opposite senses.
\end{lemma}

{\bf{Proof:}} See page 261 in \cite{hardy1952inequalities}. \hfill $\Box$

\section{Optimal solution}

\label{sec4}
For a job scheduled in the $r$th position in a sequence, $p_{[r]}$ and $a_{[r]}$ denote the actual processing time and the normal processing time of the job, respectively. All the jobs are available for processing at time zero. By Lemma~\ref{le4}, the locations of $k$ and $l$ can be calculated. Let $i$ be the position of the last job preceding the maintenance activity. If the position of the maintenance activity is before $k$ (i.e., $i<k$), then the total cost is given by 
\begin{equation}
\label{eq:ik}
 \begin{split}
								Z &=\sum_{j=1}^n(\alpha E_j+\beta T_j+\gamma d_j^{(1)}+\delta D) \\
								& = \alpha \sum_{j=1}^k (p_{[j]}+q^{(1)}-C_{[j]}) +\beta \sum_{j=l+1}^n (C_{[j]}-p_{[j]}-q^{(2)}) \\
								&+\gamma \sum_{j=1}^n (q^{(1)}+p_{[j]})+n\delta (q^{(2)}-q^{(1)}) \\
								& = \alpha \sum_{j=1}^k j p_{[j]} + \alpha i (\mu + \sigma \sum_{j=1}^i p_{[j]})+\beta \sum_{j=l+1}^n (n-j) p_{[j]} \\
									& + \gamma(n(\mu+\sigma\sum_{j=1}^i p_{[j]}) + (n+1)\sum_{j=1}^k p_{[j]} + \sum_{j=k+1}^n p_{[j]}) + n \delta\sum_{j=k+1}^l p_{[j]} \\									  &  = n \mu \gamma + \alpha i \mu + \sum_{j=1}^n \omega_j p_{[j]},\\
									\end{split}
\end{equation}
where 
\begin{equation}\label{eq:omega_ik}
\omega _j= \begin{cases}\alpha j + \alpha i \sigma + \gamma n \sigma + (n+1) \gamma & 1 \leq j \leq i \\
						\alpha j + (n+1)\gamma                                      & i <j \leq k \\
						\gamma + n \delta                                           & k < j \leq l \\
						\beta(n-j) + \gamma                                         & l < j \leq n .
						\end{cases}
\end{equation}

If $k\leq i<l$, then we have 
\begin{equation}
\label{eq:ki}
 \begin{split}
								Z & = \alpha \sum_{j=1}^k j p_{[j]} + \beta \sum_{j=l+1}^n(n-j) p_{[j]}+\gamma\left((n+1) \sum_{j=1}^k  p_{[j]} + \sum_{j=k+1}^n p_{[j]}\right) \\
									& + n\delta(\mu + \sigma \sum_{j=1}^i p_{[j]}) + n \delta \sum_{j=k+1}^l p_{[j]} \\
									&  = n \delta \mu  + \sum_{j=1}^n \omega_j p_{[j]},\\
									\end{split}
\end{equation}
where 
\begin{equation}\label{eq:omega_kil}
\omega _j= \begin{cases}\alpha j +  \gamma (n+1) + n \delta \sigma & 1 \leq j \leq k \\
						\gamma +n\delta \sigma+n\delta                                      & k <j \leq i \\
						\gamma + n \delta                                           & i < j \leq l \\
						\beta(n-j) + \gamma                                         & l < j \leq n .
						\end{cases}
\end{equation}

If $l\leq i \leq n$, then we have
\begin{equation}
\label{eq:li}
 \begin{split}
								Z & = \alpha \sum_{j=1}^k j p_{[j]} + \beta \left( \sum_{j=l+1}^n (n-j) p_{[j]} + (n-i)(\mu+\sigma \sum_{j=1}^i  p_{[j]})\right) \\
									& + \gamma \left((n+1)\sum_{j=1}^k  p_{[j]}+ \sum_{j=k+1}^n  p_{[j]}\right) +  n \delta\sum_{j=k+1}^l p_{[j]} \\									  &  = (n-i)\beta \mu  + \sum_{j=1}^n \omega_j p_{[j]},\\
									\end{split}
\end{equation}
where 
\begin{equation}\label{eq:omega_lin}
\omega _j= \begin{cases}\alpha j + \beta(n- i) \sigma +   \gamma (n+1) & 1 \leq j \leq k \\
						\beta(n-i)\sigma + \gamma + n \delta                       & k <j \leq l \\
						\beta(n-j)+ \beta(n-i)\sigma + \gamma                      & l < j \leq i \\
						\beta(n-j) + \gamma                                         & i < j \leq n .
						\end{cases}
\end{equation}
Note that $i=n$ means no maintenance activity is necessary in the schedule. Given the processing time $a_{[j]}$, the actual processing time $p_{[j]}$ of the scheduled $j$'th job can be given as follows.
\begin{equation}
\label{eq:pa}
p_{[j]}= a_{[j]} + b \sum_{t=1}^{m-1}(1+b)^{t-1} a_{[j-t]},
\end{equation} 
where $1 \leq j \leq n$, and 
\begin{equation}
m= \begin{cases} j & \text{if  }   j \leq i  \\
						j-i    & \text{if  }   j>i  .
						\end{cases}
\end{equation}

Combining \eqref{eq:ik}, \eqref{eq:ki} and \eqref{eq:li} and using \eqref{eq:pa}, we obtain
\begin{equation}
\label{eq:Z_all}
Z = M + \sum_{j=1}^n \omega_j p_{[j]} = M + \sum_{j=1}^n W_j a_{[j]},
\end{equation}
where 
\begin{equation}
M=\begin{cases} n \mu \gamma + \alpha i \mu  & i < k \\
							  n \delta \mu                 & k \leq i < l \\
						    (n-i)\beta \mu               & l \leq i \leq n,  
						\end{cases}
\end{equation}
the positional weight 
\begin{equation}
\label{eq:w}
W_j=\omega _j +b  \sum_{t=j+1}^{m'} \omega _t (1+b)^{t-j-1},
\end{equation} 
and 
\begin{equation}
m'= \begin{cases} i & \text{if  }  1\leq j \leq i  \\
						n    & \text{if  }   i<j\leq n  .
						\end{cases}
\end{equation}

From the above analysis and the rearrangement inequality (Lemma~\ref{le5}), the problem $1\mid SLK, p_j=a_j+bt, ma\mid \sum_{j=1}^n(\alpha E_j+\beta T_j+\gamma d_j^{(1)}+\delta D)$ can be solved by the following algorithm. 

{\bf Algorithm~1.}  

Step 1. By Lemma~\ref{le4}, get the values of $k=\left\lfloor \frac{n(\delta-\gamma)}{\alpha} \right\rfloor$ and $l=\left\lfloor \frac{n(\beta-\delta)}{\beta}\right\rfloor$.

Step 2. Set $i=1$.

Step 3. For $j=1,2,\ldots, n$, obtain the positional weights $W_j$ according to \eqref{eq:w}.

Step 4. Renumber the jobs in non-decreasing order of their normal processing times $a_j$. By Lemma~\ref{le5}, arrange the job with the largest normal processing time to the position with the smallest value of $W_j$, the job with the second largest normal processing time to the position with the second smallest value of $W_j$, and so forth. Then, obtain a local optimal schedule and the corresponding total cost. 

Step 5. $i=i+1$. If $i\leq n$, then go to Step 3. Otherwise, go to Step 6.

Step 6. The optimal schedule is the one with the minimum total cost.

Then we have the following result.

\begin{theorem}

Algorithm~1 solves the problem $1\mid SLK, p_j=a_j+bt, ma\mid \sum_{j=1}^n(\alpha E_j+\beta T_j+\gamma d_j^{(1)}+\delta D)$ in $O(n^2 \log n)$ time.

\end{theorem}

{\bf{Proof:}} The correction of Algorithm~1 is guaranteed by Lemmas \ref{le3} - \ref{le5}. The running time of Steps~1 and 3 is $O(n)$, and the time complexity of Step~4 is $O(n \log n)$. Since the maintenance activity can be scheduled immediately after any one of the jobs, $n$ different positions of the maintenance activity must be considered and evaluated to obtain the global optimal solution until Step~6. Hence, the time complexity for solving the $1\mid SLK, p_j=a_j+bt, ma\mid \sum_{j=1}^n(\alpha E_j+\beta T_j+\gamma d_j^{(1)}+\delta D)$ problem is $O(n^2 \log n)$.           \hfill $\Box$

\section{Numerical example}
\label{sec5}

In this section, Algorithm~1 is demonstrated by the following example.

{\bf Example~1.} There are $n=9$ jobs. The normal processing times of jobs are $a_1=62$, $a_2=81$, $a_3=25$, $a_4=82$, $a_5=26$, $a_6=19$, $a_7=55$, $a_8=9$ and $a_9=91$. Let the common deteriorating factor $b=0.05$. The penalties for unit earliness, tardiness, due-window starting time and due-window size are $\alpha =4$, $\beta =15$, $\gamma =5$ and $\delta =6$, respectively. The basic maintenance time is $\mu = 10$ and the deteriorating maintenance factor is $\sigma=0.1$.

\textit{Solution:} By Lemma~\ref{le4}, we have the locations of $k=\left\lfloor \frac{n(\delta-\gamma)}{\alpha} \right\rfloor = 2$ and $l=\left\lfloor \frac{n(\beta-\delta)}{\beta}\right\rfloor=5$. 
In the following we explain the cases $i=1$, $i=3$ and $i=6$ and the other cases are similar. If $i=1$, the maintenance time is immediately after the first job is finished. Since $i < k$, according to \eqref{eq:omega_ik}, we have $\omega_1=58.9$, $\omega_2=58.0$, $\omega_3=59.0$, $\omega_4=59.0$, $\omega_5=59.0$, $\omega_6=50.0$, $\omega_7=35.0$, $\omega_8=20.0$ and $\omega_9=5.0$, and we obtain $W_1=58.9000$, $W_2=73.9324$, $W_3=71.3642$, $W_4=67.9659$, $W_5=64.7294$, $W_6=53.0756$, $W_7=36.2625$, $W_8=20.2500$ and  $W_9=5.0000$. Based on Algorithm~1, we have the local optimal sequence is (7,8,6,3,5,1,2,4,9) and the total cost is $Z=17476.37$. If $i=3$, the maintenance time is immediately after the third job is finished. Since $k < i$, according to \eqref{eq:omega_kil}, we have $\omega_1=59.4$, $\omega_2=63.4$, $\omega_3=64.4$, $\omega_4=59.0$, $\omega_5=59.0$, $\omega_6=50.0$, $\omega_7=35.0$, $\omega_8=20.0$ and $\omega_9=5.0$, and we obtain $W_1=65.9510$, $W_2=66.6200$, $W_3=64.40000$, $W_4=67.9659$, $W_5=64.7294$, $W_6=53.0756$, $W_7=36.2625$, $W_8=20.2500$ and  $W_9=5.0000$. Based on Algorithm~1, we have the local optimal sequence is (3,6,7,8,5,1,2,4,9) and the total cost is $Z=17634.66$. If $i=6$, the maintenance time is immediately after the sixth job is finished. Since $l < i$, according to \eqref{eq:omega_lin}, we have $\omega_1=58.5$, $\omega_2=62.5$, $\omega_3=63.5$, $\omega_4=63.5$, $\omega_5=63.5$, $\omega_6=54.5$, $\omega_7=35.0$, $\omega_8=20.0$ and $\omega_9=5.0$, and we obtain $W_1=75.4469$, $W_2=75.6637$, $W_3=73.0131$, $W_4=69.5362$, $W_5=66.2250$, $W_6=54.5000$, $W_7=36.2625$, $W_8=20.2500$ and  $W_9=5.0000$. Based on Algorithm~1, we have the local optimal sequence is (6,8,3,5,7,1,2,4,9) and the total cost is $Z=18271.87$.

\begin{table}[!htb]
	\centering
	\caption{The corresponding local optimal job sequences and total costs with one maintenance activity at all possible positions in Example~1.}
	\label{tbl1}
	\begin{tabular}{ccc}
		\hline
		$i$ & Job sequence & $Z$ \\ \hline
		1 & (7,8,6,3,5,1,2,4,9) & \underline{17476.37} \\
		2 & (7,5,8,6,3,1,2,4,9) & 17525.07 \\
		3 & (3,6,7,8,5,1,2,4,9) & 17634.66 \\
		4 & (6,8,3,7,5,1,2,4,9) & 17749.44 \\
		5 & (6,8,3,5,7,1,2,4,9) & 18157.92 \\
		6 & (6,8,3,5,7,1,2,4,9) & 18271.87 \\
		7 & (6,8,3,5,7,1,2,4,9) & 18347.63 \\
		8 & (6,8,3,5,7,1,2,4,9) & 18170.85 \\
		9 & (6,8,3,5,7,1,2,4,9) & 17519.13 \\ 
		\hline 
	\end{tabular}
\end{table}

As shown in Table~\ref{tbl1}, all the local optimal job sequences and the corresponding total costs are presented, among which the optimal total cost is underlined. The global optimal solution for this example includes the following: (i) the job sequence is (7,8,6,3,5,1,2,4,9) and the corresponding job  starting time and actual processing time are (0.00, 70.50, 79.50, 98.95, 125.37, 154.12, 220.30, 308.79, 402.70) and (55.00, 9.00, 19.45, 26.42, 28.74, 66.18, 88.49, 93.91, 107.61), respectively; (ii) the slack window parameters are $q^{(1)}=79.50
$ and $q^{(2)}=154.12$; (iii) the maintenance activity is located immediately after the first job (i.e. Job 7), starting at time $t=55.00$
 and ending at time $t=70.50$; (iv) the total cost is $Z=17476.37$.

\section{Conclusion}
\label{sec6}

We solved a single machine slack due-window assignment and scheduling problem of linear time-dependent deteriorating jobs and a deteriorating maintenance activity, and gave a polynomial-time algorithm. The running time of this algorithm does not exceed $O(n^2logn)$. Further research may consider the problem with the setting of parallel identical machines, or the problems with min-max type objective functions.

\end{document}